\documentclass[11pt]{iopart} 
\usepackage{graphicx}
\usepackage{amsmathred}
\usepackage{color}
\usepackage{cancel}
\usepackage{soul}
\usepackage{ulem} %to strike out
\usepackage{marginnote}
\usepackage{amsfonts}
\usepackage{latexsym}
\usepackage{relsize}
\usepackage{euscript}%changes caligraphic font, see http://www.maths.usyd.edu.au/u/SMS/texdoc/euscript.pdf
\usepackage{amsbsy}
\usepackage{dsfont}
\usepackage{hyperref}
\usepackage{amsthm}

\newcommand{\da}[0]{\dagger}

\newcommand{\be}{\begin{equation}}
\newcommand{\ee}{\end{equation}}
\newcommand{\bea}{\begin{eqnarray}}
\newcommand{\eea}{\end{eqnarray}}

\newcommand{\pa}{\partial}

\begin{document}

\title[Effects of gravity and motion on quantum entanglement in space-based experiments]{\sf \bfseries Testing the effects of gravity and motion on quantum entanglement in space-based experiments}

\author{\sf \bfseries  David Edward Bruschi$^{1}$,Carlos Sab{\'\i}n$^{2}$, Angela White $^{3}$, Valentina Baccetti $^{4}$, Daniel K. L. Oi $^{5}$, Ivette Fuentes$^{2}$ }

\address{
$^{1}$School of Electronic and Electrical Engineering, University of Leeds, Woodhouse Lane,  Leeds, LS2 9JT,  United Kingdom//$^{2}$School of Mathematical Sciences, University of Nottingham, University Park,
Nottingham NG7 2RD, United Kingdom\\ $^{3}$ Joint Quantum Centre (JQC) Durham-Newcastle, School of Mathematics and Statistics, Newcastle University, Newcastle upon Tyne, NE1 7RU, United Kingdom\\$^{4}$School of Mathematics, Statistics and Operations Research, Victoria University of Wellington, PO Box 600, Wellington 6140, New Zealand\\$^{5}$SUPA Department of Physics, University of Strathclyde, Glasgow G4 0NG,United Kingdom}
\ead{david.edward.bruschi@gmail.com}

\begin{abstract}
We propose an experiment to test the effects of gravity and acceleration on quantum entanglement in space-based setups. We show that the entanglement between excitations of two Bose-Einstein condensates is degraded after one of them undergoes a change in the gravitational field strength. This prediction can be tested if the condensates are initially entangled in two separate satellites while being in the same orbit and then one of them moves to a different orbit. We show that the effect is observable in a typical orbital manoeuvre of nanosatellites like CanX4 and CanX5. 
\end{abstract}
\maketitle
\section[1]{Introduction}
Quantum mechanics and relativity are the two most fundamental theories of the Universe known to science. Despite both working extremely well in predicting and quantifying effects in their respective regimes of application, they are commonly deemed as incompatible. On one hand, quantum mechanics predicts with great accuracy the behaviour of microscopic particles that can be in asuperposition of being in two different places at once. On the other hand, general relativity provides an effective description of the Universe at large length scales where time can flow at different rates in different places. However, we do not fully understand what happens when these effects occur together. The inability to unify these theories remains one of the biggest challenges in theoretical physics today.
\begin{figure}[t!]
\includegraphics[width=\linewidth]{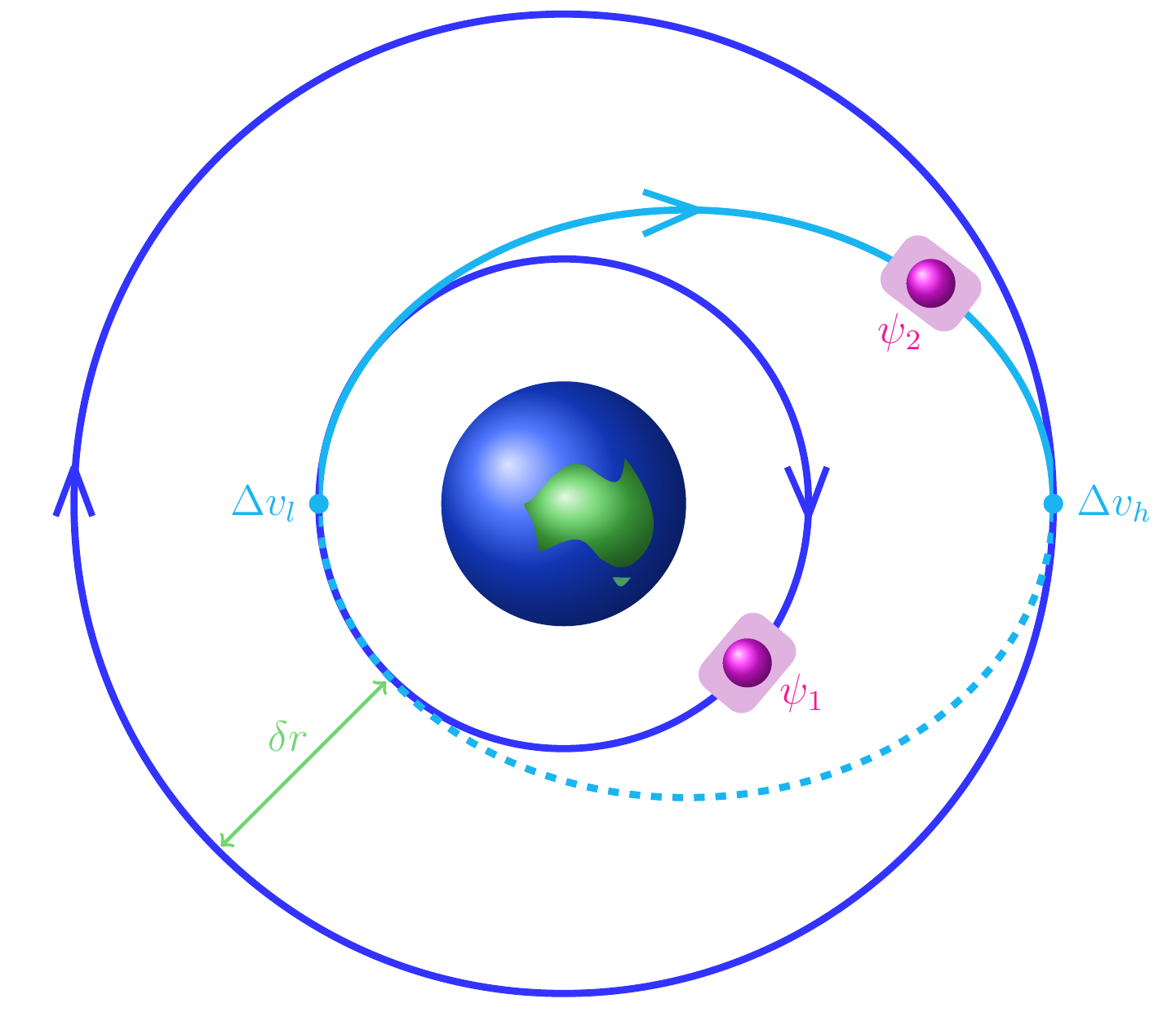}
\caption{Experimental proposal. Two BECs inside separate satellites are entangled while both are in the same circular LEO orbit. Then one of them undergoes acceleration during a finite time in order to change to a different circular orbit, by means of a Hohmann transfer obit.} \label{fig:sketch}
\end{figure}

Understanding general relativity at small length scales where quantum effects become relevant is a highly non-trivial endeavour that has suffered from a lack of experimental guidance. An alternative approach is to study quantum effects at large scales, which promises to be experimentally achievable in the near future \cite{rideout, ursin}. Cutting-edge quantum experiments are reaching relativistic regimes, where the effects of gravity and motion on quantum properties can be experimentally tested.  In 2012 a teleportation protocol was successfully performed across 144 $\operatorname{km}$ by the group lead by A. Zeilinger \cite{zeilingerteleport}. Motivated by this success and related experimental developments \cite{panteleport, plane,daniel}, major space agencies, e.g. in Europe and Canada, have invested resources for the implementation of space-based quantum technologies \cite{spaceexp1,spaceexp2,spaceexp3}. There are advanced plans to use satellites to distribute entanglement for quantum cryptography and teleportation (e.g. Space-QUEST project) and to install quantum clocks in space (e.g. Space Optical Clock project). Such experiments are of great interest since relativistic effects can be expected at the regimes where satellites operate. For instance, it is well-known that the Global Positioning System (GPS), a system of satellites used for time dissemination and navigation, requires relativistic corrections to determine time and positions accurately. Indeed previous theoretical work has addressed these fundamental questions by showing that gravity, motion and space-time dynamics can create and degrade entanglement \cite{ivyalsingreview}. For instance, recent work \cite{teleportationico} shows that acceleration produces observable effects on quantum teleportation. However, current experimental space-based designs are yet to consider these findings. In this paper we propose a space-based experiment to test the effects of gravity and motion on quantum entanglement.
\

Most proposals to implement quantum technologies in space have been developed within the framework of quantum mechanics \cite{onofrio}. However, quantum mechanics is a non-relativistic theory where the effects of acceleration and gravity can only be added ad-hoc.  
The correct arena in which to look for relativistic effects is Quantum Field Theory (QFT), which describes the behaviour of quantum fields in space-time. It is a semiclassical description in the sense that mater and radiation are quantised but the space-time is treated as a classical background. However, unlike quantum mechanics, QFT naturally incorporates Lorentz invariance, as required by the postulates of relativity theory. Indeed, QFT successfully merges quantum theory and special relativity in the framework of the standard model of elementary particles.  Moreover,  QFT in curved spacetime provides some answers to questions about the overlap of quantum mechanics and general relativity \cite{birrelldavies}. Very recently we have started to see some of its predictions be experimentally verified \cite{casimirwilson, casimirwestbrook, facciohawking}. An appropriate QFT approach that includes the effects of relativity on entanglement has been the main ingredient missing in current proposals to use entanglement in space-based implementations of quantum technologies \cite{rideout, ursin}. These ideas have also been discussed by Downes and Ralph \cite{tonytim} who have pointed out that in order to correctly account for effects that take place at increasing length and shorter time scales, quantum information must be extended to a fully relativistic setting.

 In this paper we use a QFT framework to show that the gravitational field of the earth and accelerated motion can induce experimentally observable effects on the basic resource for quantum information and communication tasks, namely quantum entanglement. Our findings, on one hand, shed light on fundamental questions about the overlap of quantum theory and relativity and, on the other hand, will enable experimentalists to correct negative gravitational effects on quantum information. Our research program aims, not only to characterise relativistic effects so that they can be corrected, but also to learn how to exploit them in order to improve the performance of quantum technologies in space. 

Recently, it has been shown that the entanglement between field modes of localised systems, such as cavities, is sensitive to changes in acceleration \cite{alphacentauri}.  Via the equivalence principle, this means that entanglement should, therefore, be affected by changes in gravitational field strengths.  We propose to demonstrate this experimentally by considering the entanglement between the excitations of two Bose-Einstein Condensates (BECs), each one of them prepared in a separate satellite.  The BEC excitations we consider are known as quasiparticles or phonons. These excitations obey, under certain circumstances, a massless Klein-Gordon equation with a very slow speed of propagation  \cite{pethicksmith}. Low propagation speeds is the key element to enable the observation of the effect we describe below within realistic experimental regimes. We propose to entangle two BEC modes, one in each BEC, while the BECs move close to each other along the same circular earth orbit. One of the satellites will then undergo non-uniform motion to change to an orbit subject to a different gravitational field strength (see Fig. (\ref{fig:sketch})).  Our analysis shows that the entanglement degradation between the BEC modes is a periodic function of the change in gravitational field strength in the orbit. This effect is significant already for typical parameters involved in microsatellite manoeuvres, which is a great advantage since experiments involving such satellites have relatively low costs.

\section[2]{Model and results}
Let us explain our methods and results in more detail. In the absence of atomic collisions, a BEC can in principle reach absolute zero temperature and be described by a classical mean field. However, collisions are always present and therefore, in the superfluid regime, the condensate is better described by a mean field classical background plus quantum fluctuations. The fluctuations, for length scales larger than the so-called healing length, behave like a phononic quantum field.  The classical background energy density, pressure and number density play the role of an effective spacetime metric which in principle can be curved. The dependence of this metric on the BEC parameters will be presented below. The field $\Pi(\xi)$ can be expanded in terms of the so-called Bogoliubov modes $\phi(\xi)$ \cite{pethicksmith},

\begin{align}
 Ê Ê\Pi(\xi)Ê Ê&= Ê\sum\limits_{k}\,\Bigl(\,\phi_{k}(\xi)\,a_{k}\,+\,\phi^{*}_{k}(\xi)\,a^{\dagger}_{k}\Bigr)\,.
 Ê Ê\label{eq:scalar field in-region}
\end{align}
We use $\xi$ to denote arbitrary coordinates. The operators $a_k$ and $a^{\dagger}_k$ associated with the modes are annihilation and creation operators, respectively,  which obey the standard  canonical commutation relations. The dispersion relation is given by $\omega_k=c_s\,|\bf{k}|$ where $c_s$ is the speed of sound.

In a homogenous condensate, the modes obey a massless Klein-Gordon equation $\Box\Pi=0$ where the d' Alembertian operator $\Box=1/\sqrt{-\mathfrak{g}}\,\partial_{a}(\sqrt{-\mathfrak{g}}\mathfrak{g}^{ab}\partial_{b})$ depends on an effective spacetime metric $\mathfrak{g}_{ab}$ -with determinant $\mathfrak{g}$- given by (see Appendix A): \cite{Visser:2010xv,liberati}

\begin{equation}
\mathfrak{g}_{ab}=\left(\frac{n^2_0\,c_s^{-1}}{\rho_0+p_0}\right)\left[g_{ab}+\left(1-\frac{c_s^2}{c^2}\right)V_aV_b\right].
\end{equation}
Note that this metric is a function of background mean field properties of the BEC such as the number density $n_0$, the energy density $ \rho_0$ and the pressure $p_0$. The effective curvature naturally arises from decoupling the field equations of the background mean field and the quantum fluctuations.  $V_a$ is the BEC 4-velocity with respect to the laboratory reference frame, while $g_{ab}$ is the background, real spacetime metric that in general may be curved.  Strictly speaking, in the experiment we propose, the BECs move in a Schwarzschild metric.  However, due to the smallness of the Schwarzschild radius of the earth, it is reasonable to assume that the spacetime is flat.  The BECs are inertial while they free fall in a circular orbit, and in this case we use Minkowski coordinates $(t,\vec{x})$.  In order to change the orbit of one of them, so that it undergoes a change in gravitational potential, acceleration is required.  We consider that the satellite undergoes a single change in velocity, that is a single period of uniformly accelerated motion. The direction, intensity and duration uniquely determines the new orbit. 
Therefore, we consider a  Rindler transformation of the Bogoliubov modes  since Rindler coordinates are suitable to describe periods of uniformly accelerated motion (see Appendix B).  We choose the comoving frame $V_a=(c;0,0,0)$ since we want to describe the effects in the rest frame of the BEC. Under these conditions we obtain an effective metric $\mathfrak{g}_{ab}$ which is also conformally flat (see Appendix A).

For the sake of simplicity, we consider a quasi one dimensional BEC.  Suitable close to hard-wall boundary conditions \cite{condensatebox1, condensatebox2,condensatebox3} allow us to consider a spectrum similar to the well-known spectrum of an optical cavity  given by $\omega_n=2\pi\times \frac{n\,c_s}{L}$,  where $L$ is the length of the cylinder.
Initially two space experimentalists, Valentina and Yuri, prepare a two-mode squeezed state between two inertial BECs  with squeezing parameter $r>0$. Details on how to prepare such state are discussed in section C. The quantum correlations of this state are fully characterised by the reduced covariance matrix of the two modes $\sigma_{kk'}$, a real $4\times4$ matrix that only depends on $r$ (see Appendix C).  In particular, entanglement can be quantified with the negativity which for this state is given by \cite{gerardomeasurements} 
\begin{equation}
N^{(0)}=\operatorname{max}[0,\frac{1}{2} (e^{2r}-1)].\label{eq:initialneg}
\end{equation}
where the condensate undergoes free evolution.
After preparing the initial state Yuri moves his BEC into an orbit subject to a different gravitational potential. This is achieved by accelerating with constant acceleration $a$ for a proper time $\tau$ as measured by an observer at the center of the rigid trap \cite{synge}. Once in the new orbit the BEC moves inertially again. It is important to make sure that the motion preserves the homogeneity of the condensate. In principle, the acceleration can make the condensate become inhomogeneous. However, these effects are negligible in the regimes considered in our discussion (see Appendix B). The initial covariance matrix $\sigma$  of the state including all modes is transformed after the change in orbit into $\tilde{\sigma}=S\sigma S^{T}$ where $S$ is a symplectic matrix that encodes the time evolution of the system. The reduced covariance matrix $\tilde{\sigma}_{kk'}$ for the two particular modes $k$ and $k'$ of interest can be obtained from $\tilde{\sigma}$.
During inertial and uniformly accelerated segments of motion, the field modes only undergo free evolution. Therefore, the transformation in this case is simply composed of local rotations with angles $\omega_{k}t$ and $\omega_{k'}t$, where $\omega_{k}$ and $\omega_{k'}$ are the angular frequencies of the modes $k$ and $k'$ respectively.  However, during changes from inertial to accelerated motion, the modes undergo a  Bogoliubov transformation with coefficients ${}_0\alpha_{mn}$ and ${}_0\beta_{mn}$ that relate the mode functions in the Minkowski and Rindler frames~\cite{nicoivy}. The coefficients ${}_0\alpha_{mn}$ account for mode mixing within the moving condensate, while ${}_0\beta_{mn}$ account for particle pair production. Therefore, the total Bogoliubov coefficients $\alpha_{mn}$, $\beta_{mn}$ are functions of  ${}_0\alpha_{mn}$, ${}_0\beta_{mn}$ and of phases acquired during the period of uniform acceleration. They can be computed analytically (see Appendix C) using a perturbative expansion in the parameter:
\begin{equation}
h=%\frac{a\,L}{c^{2}}\ll 1\,,
a\,L/c_s^{2}\ll 1\,.
\label{eq:pertparam}
\end{equation}
 We can write the coefficients as $\alpha_{mn}=\alpha^{_{(0)}}_{mn}+\alpha^{_{(1)}}_{mn}h+O(h^{2})$ and $\beta_{mn}=\beta^{_{(1)}}_{mn}h+\beta^{_{(2)}}_{mn}h^{2}+O(h^{3})$. After the change of orbit, we find that the entanglement has changed and is now given by $N=N^{(0)}+N^{(2)}\,h^2+ \mathcal{O}(h^4)$. More specifically:
 \begin{equation}
 N=\operatorname{max}[0,N^{(0)}(1-e^{2r}(f^{\alpha}_{k'}+f^{\beta}_{k'})\,h^2)-e^{2r}f^{\beta}_{k'}\,h^2] \label{eq:neg}
 \end{equation} 
 where $f^{\alpha}_{k'}$, $f^{\beta}_{k'}$ are functions of the Bogoliubov coefficients that depend periodically on the difference of gravitational field strength (see Appendix C).  Note that $N^{(0)}$ is the entanglement of the initial state given by Eq. (\ref{eq:initialneg}). $N$ is always smaller than $N^{(0)}$, since the entanglement is degraded by mode mixing and particle creation \cite{casimirwilson,teleportationico}. This degradation effect becomes observable for large enough, but still perturbative, values of $h$, $h^2\simeq0.05$ \cite{teleportationico}. In optical cavities, these values of $h$ are obtained with accelerations of $10^{23}\,m/s^2$ -see Eq. (\ref{eq:pertparam})- while in superconducting cavities, the corresponding order of magnitude is $10^{17}\,m/s^2$-which can be achieved by non-mechanical means \cite{casimirwilson,teleportationico}. In the case under study here namely,  BECs, the typical values $L\simeq100\,\operatorname{\mu\,m}$ and $c_s=1\,\operatorname{mm/s}$ give rise to $a\simeq10^{-3}\,m/s^2$.
  
  \begin{figure}[t!]
\includegraphics[width=\linewidth]{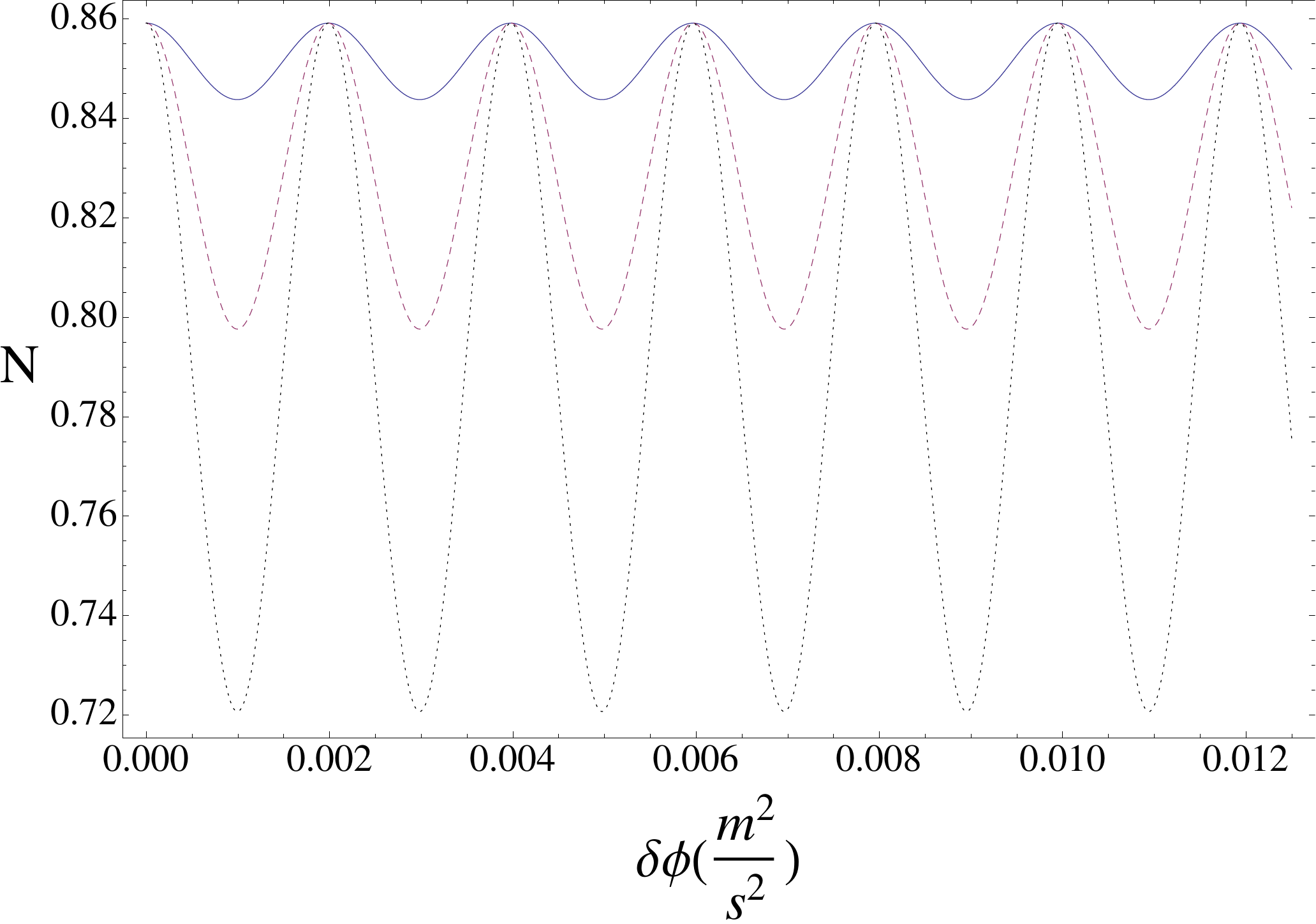}
\caption{Negativity $N$ vs. difference in gravitational field strength between initial and final orbits $\delta \phi$, after the first change in velocity $\Delta v_l$. The acceleration of the satellite is $a=10^{-3} \operatorname{m/s^2}$ (solid, blue), $a=2\,\cdot 10^{-3} \operatorname{m/s^2}$ (red,dashed), $a=3\, \cdot10^{-3} \operatorname{m/s^2}$ (black, dotted) while $L=100\,\operatorname{\mu\,m}$, $c_s=1 \operatorname{mm/s}$, giving rise to $h^2\simeq 0.05$ and $\Omega_1=2\pi\times 50\,\operatorname{Hz}$. The initial squeezing is $r=1/2$. }\label{fig:results}
\end{figure}

\section[3]{Experimental setup}
We now assess the feasibility of testing the degradation of entanglement due to orbit changes with a space-based experiment using a pair of nanosatellites. Nanosatellites are fully functional spacecraft with a mass of 1 to 10kg. The use of conventional off the shelf (COTS) parts, component miniaturization, and standardized systems means that they are a comparatively low cost avenue to space. Capabilities such as power, attitude and position control, propulsion, optics, communication, and autonomous operation are under active development which greatly expands the missions which may be undertaken within the mass and volume envelope of the nanosatellite platform. At the same time, quantum experiments have also become more compact which makes it feasible to place them on small satellites~\cite{daniel}.

An example of the capability required for such an experiments is the pair of CanX-4 and CanX-5 \cite{cubesats,cubesats2} satellites due to launch in 2013. These are built according to the Generic Nanosatellite Bus (GNB) specification which consists of a 20cm a side cube with a mass of approximately 7.5kg. Typically, such a spacecraft will have a mission payload volume of 1.8 litres and mass of 2kg. The CanX-4/5 pair will demonstrate formation flying in orbit and are each equipped with high precision differential GPS receivers for ~cm relative positioning determination, and a single axis thruster allowing orbit changes. The latter consists of the Canadian Nanosatellite Advanced Propulsion System (CNAPS) and has a rated thrust of $20\, \operatorname {mN}$ and an $I_{sp}$ of $35 \, \operatorname{s}$ resulting in a $\Delta\, V$ of $11.1 \operatorname{m/s}$. Therefore the satellites can accelerate with the constant acceleration $a\simeq10^{-3} \operatorname{m/s^2}$ necessary to make the predicted effects observable. 
Let us consider a pair of satellites, such as CanX4 and CanX5, moving along the same circular orbit. Each satellite contains a BEC with initially entangled phonon modes. Such an entangled state can be prepared in several ways. For instance, the BECs can be made to interact through Bragg scattering with two separated laser beams  that excite quasi-particles of specific momenta in each condensate. Entanglement is then produced by performing projective measurements on the scattered light beams \cite{Deb:2008}. Atom-light entangling techniques can also be used, where via  electromagnetically induced transparency and subsequent projective measurements, the entanglement is transferred from two probe laser beams to two spatially separated BECs \cite{Kuang:2007}. Similar techniques can also be applied by considering two separate BECs in two distinct, high-finesse optical cavities, on which two quantum correlated light fields are incident, hence transferring their quantum correlated state to the two BECs, \cite{Kumar:2011}. If the BEC is in an initial thermal state instead of the vacuum state, the amounts of initial squeezing and entanglement that can be generated will be lower \cite{davidnicoivy}. In order to generate a squeezing of $r=1/2$ at frequencies of  $100 \, \operatorname{Hz}$, the BEC should be cooled down to a few $\operatorname{nK}$.
Finally, notice that  the  experimental setup required to create and hold the BEC can be as small as 0.5 $\operatorname{L}$ \cite{portablebec}. Important efforts are currently taking place to load and maintain a BEC on a chip device in space. See, for example, the QUANTUS project \cite{quantus} aimed at using a BEC to detect microgravity effects in space.

The effects predicted in this work arise when a satellite undergoes a change of circular orbit, determined by the difference in gravitational field strength between the initial and final orbits. As an example, the change of orbit can be achieved in an efficient and elegant manner by means of a Hohmann transfer orbit \cite{hohhmann, hohhmann1} (see Fig.1). The procedure is the following. First a change of velocity $\Delta\,v_l$  moves the satellite to an elliptic orbit.  Then the satellite navigates half of this new orbit,  before finally a second velocity kick $\Delta\,v_h$ puts the satellite back into a circular orbit. The difference  between the radius of the initial orbit $r_l$ and the radius of the final orbit $r_h$ determines the magnitude of the velocity kicks through the relations
\begin{align}
\Delta v_l=&\sqrt{\frac{GM}{r_l}}\left(\sqrt{\frac{2r_h}{r_l+r_h}}-1\right)\nonumber\\
\Delta v_h=&\sqrt{\frac{GM}{r_h}}\left(1-\sqrt{\frac{2r_l}{r_l+r_h}}\right),\label{eq:kicks}
\end{align}
where $G$ is Newton's gravitational constant and $M$ is the mass of the earth. In particular, assuming a small change of altitude $r_h=r_l+\delta\, r$ with $\delta\,r<< r_l$, we find $\Delta\,v_l\simeq \Delta\,v_h\simeq\sqrt{\frac{GM}{r_h}}\frac{\delta\,r}{4\,r_h}\simeq\sqrt{\frac{r_h}{GM}}\frac{\delta\,\phi}{4}\simeq 3\, 10^{-3} \operatorname{m/s}$ for a Low Earth Orbit (LEO) of $r_h=R_e+400\, km$ -$R_e$ being the radius of the earth and $\delta\, \phi$ the difference in gravitational field strength between the initial and final orbits. Therefore, for constant acceleration, each radial distance between circular orbits is related with a different duration of the acceleration.  The whole manoeuvre takes half-period $P/2$ of the elliptical transfer orbit $P\simeq2\pi\sqrt{r_h^3/GM}\simeq 5000\,s$ which is larger than the average lifetime of a BEC. However, the degradation of the entanglement takes place immediately after the first change in velocity, and can be observed during the navigation of the transfer orbit. Eqs. (\ref{eq:neg}) and (\ref{eq:kicks}) imply that the entanglement oscillates with the radial distance between the initial and final orbit, or equivalently, with the difference in the gravitational strength. In Fig. (\ref{fig:results}) we show that, for realistic experimental parameters, oscillations have a significant amplitude and a period of around 2 $m$, meaning that almost any change of orbit would lead to an observable effect on the initial quantum entanglement. Note that the duration of the acceleration in the plot is of the order of $0.1 \operatorname{s}$. The maximum change of velocity is $\Delta v_l\simeq10^-3\,\operatorname{m/s}$ well within reach of current technologies since CanX4 and CanX5 are capable of achieving maximum changes of velocities of $\Delta v=11.1\, \operatorname{m/s}$. Much larger changes of orbit can be considered for which the behaviour of entanglement as a function of difference in gravitational strength is shown in Fig. (\ref{fig:results}). Since CanX4 and CanX5 are designed to determine positions with an accuracy of $\operatorname{cm}$, they seem ideal devices to analyse the dependence of entanglement with the radial distance. 

The readout of the quantum correlations might be performed in a  manner similar to the experiment in \cite{casimirwestbrook}, where upon releasing the condensate trapping potential, each phonon is converted into an atom with the same momentum and velocities are measured by a position sensitive single-atom detector. Unfortunately this technique is destructive and many shots of the experiment would be necessary to achieve the required statistics. An alternative method consists in using atomic quantum dots or optical lattices coupled to each condensate to probe the reduced field states of each condensate \cite{nuestrotermo}. This method enables one to perform several thousands of correlated measurements within the coherence time of the entangled state we consider. For weakly dissipative systems the coherence time is given by $t=\hbar/(mc^2)$ \cite{buschparentani}. Considering that the speed of sounds is $c_s=1 \operatorname{mm/s}$ (as shown in Fig. \ref{fig:results}) and the mass of ${}^{4} He$ is four times the mass of the proton, we obtain that $t\simeq100 \operatorname{ms}$. On the other hand, the interaction between each dot and the condensate can be modulated through Feshbach resonances in the sub-ms regime \cite{feshbachlucia} and a number of 1500 dots can be considered \cite{nuestrotermo}. This results in the possibility of making $10^5$ measurements in 100 $\operatorname{ms}$. An alternative method to measure the covariance matrix of a pair of phononic modes through non-destructive measurements has been recently introduced in \cite{carusottofinazzi}.
The detection of quantum entanglement between phononic modes in BECs is currently a topic of great interest  \cite{davidnicoivy,buschparentani,carusottofinazzi}. Important steps in this direction have already been given in \cite{casimirwestbrook,megamind}. In particular, in \cite{megamind} the authors measure quantum fluctuations of the number of phonons in a particular mode,  by using in-situ techniques. Since in our case entanglement is proportional to the squeezing parameter, a measurement of two-mode squeezing would also be an indirect estimator of the predicted effects. Given the accelerated rate at which state-of-the-art experiments in BECs take place, it is foreseeable that it will be possible to detect quantum correlations between phonon modes in the near future. The degradation effect that we predict can be as large as 20 $\%$ (see Fig. \ref{fig:results}) of the initial entanglement and has a characteristic dependence on the magnitude and duration of the acceleration. 

\section[4]{Conclusions} 

In conclusion, we have shown that changes in the gravitational field strength produce effects on quantum entanglement that are observable in space-based experiments. In particular, we have shown that entanglement between two BECs inside separate satellites can be degraded when one of them undergoes a change of orbit. Entanglement oscillates periodically with the difference in gravitational potential of the orbits.  Therefore, by accurately controlling  the satellite positions, it is possible to find situation in which entanglement is conserved. Our results shed light on fundamental aspects in the overlap between quantum theory and relativity by working within QFT, a framework which incorporates appropriately these theories in regimes where satellites operate. These results will inform future space-based quantum technologies, including quantum key distribution and other quantum cryptographic experiments. A comprehensive understanding of relativistic effects on quantum properties will enable us not only to make the necessary corrections to the technologies they affect, but also opens up the possibility of using relativistic effects as resources.

In honour of Valentina Tereshkova and Yuri Gagarin who where the first woman and man to go to space.
\subparagraph{Appendices}
\subparagraph{Appendix A: Description of a Bose-Einstein condensate on an underlying spacetime}
The Lagrangian density of a Bose-Einstein condensate on a spacetime metric $g_{ab}$ trapped by an external potential $V(x^{\mu})$ is given by \cite{liberati},
\begin{equation}
\label{rel_bec}
\mathcal{\hat{L}}=\sqrt{-g}\,g^{ab}\pa_a \Phi^\dagger \pa_b \Phi -\left(\frac{m^2c^2}{\hbar^2} +V(x^{\mu})\right)
\Phi^\da \Phi-U(\Phi^\da\Phi;\lambda_i).
\end{equation}
where $c$ is the speed of light, $\hbar$ Planck's constant and $g=\det{g_{ab}}$.  The atomic field $\Phi$ consists of $N$ atoms of mass $m$ that interact with each other through $U(\hat{\phi}^\da\hat{\phi};\lambda_i)$.
The interaction strengths $\lambda_i$ can in principle depend on the coordinates $x^{\mu}$ of the background space-time. In the regime below the critical temperature $T_c$, the atomic field can be approximated by $\Phi=\Phi_0(1+\Pi)$, where  $\Phi_0$ is a classical background field and $\Pi$ is a quantum field corresponding to fluctuations known as phonons. In this regime, the background field obeys the non-linear Klein-Gordon equation
\begin{equation}
\label{eqmot}
 \Box_{g} \Phi_0-\left(\frac{m^2 c^2}{\hbar^2}+V(x^{\mu})\right)\Phi_0-U'(\rho;\lambda_i)\Phi_0=0
\end{equation}
where $\rho:=\Phi_0^*\Phi_0$ is the background density and $\Box_{g}:=\sqrt{-g}^{-1}\partial_{a}(\sqrt{-g}\partial^{a})$ is the d'Alambertian operator. The superscript in $U'$ denotes the derivatives with respect to $\rho$. Eq.\eqref{eqmot} reduces to the standard Gross-Pitaevskii equation in the Newtonian limit $c^2\rightarrow \infty$ \cite{liberati}. On the other hand, the quantum fluctuations $\Pi$ obey the field equation
\begin{equation}
\label{pert}
\Box_g \Pi+2 g^{ab}\left(\pa_a \ln \Phi_0 \right)\pa_b \Pi-\rho U''(\rho;\lambda_i)=0.
\end{equation}
Writing $\Phi_0=\sqrt{\rho}e^{i\theta}$ we define the generalized kinetic operators as $T_\rho \equiv -\frac{\hbar^2}{2m}\left(\Box_g+g^{ab}\pa_a \ln \rho \, \pa_b\right)$, the effective speed of phonon propagation $c^2_0 \equiv \frac{\hbar^2}{2 m^2} \rho U''(\rho; \lambda_i)$ and the four velocity vectors $u^{a} \equiv \frac{\hbar}{m} g^{ab}\pa_b \theta$. We can then rewrite the equation as,
\begin{equation}
\left\{ \left[ i \hbar u^a \pa_a + T^\rho\right]\frac{1}{c^2_0}\left[-i \hbar u^b \pa_b+ T^\rho \right] -\frac{\hbar^2}{\rho} g^{ab}\pa_a \rho \pa_b \right\}\Pi=0,\label{almost:final}
\end{equation}
$T_\rho$ can be neglected when the dispersion relation for the perturbations is $\omega^2_{-}=c^2_s k^2$ and in the eikonal approximation \cite{liberati}. That is when the background quantities vary slowly in space and time on scales comparable with the wavelength and the period of the perturbations, respectively \cite{liberati}. This assumption is equivalent to neglecting the quantum pressure term in the Gross-Pitaevskii equation obtained in the Newtonian limit.
In this case Eq. \eqref{almost:final} becomes the Klein-Gordon equation
\begin{equation}
\Box_{\mathfrak{g}}\Pi=\frac{1}{\sqrt{-\mathfrak{g}}}\pa_a (\sqrt{-\mathfrak{g}}\, \pa^a \Pi)=0, \label{klein:gordon:equation}
\end{equation}
where the effective metric $\mathfrak{g}_{ab}$ is defined as
\begin{equation}
\label{metric}
\mathfrak{g}_{ab}=\frac{\rho}{\sqrt{1-u_d u^d/c^2_0}}\left[g_{ab}\left(1-\frac{u_d u^d}{c^2_0}\right)+\left(\frac{u_a u_b}{c^2_0}\right) \right]
\end{equation} 
By defining the four-velocity $v^a\equiv \frac{c}{\| u \|} u^a$ and the scalar speed of sound $c^2_s = \frac{c^2 c^2_0/ \|u\|^2}{1+c^2_0/ \|u\|^2}$, the effective metric can be written as
\begin{equation}
\ \mathfrak{g_{ab}}=\frac{c}{c_s}\frac{n_0}{\varrho_0+p_0} \left[g_{ab}+\left( 1- \frac{c_s^2}{c^2}\right)v_a v_b \right]\label{metric:final}
\end{equation}
The conformal factor in the last equation \eqref{metric:final} can be found by considering the hydrodynamical description for a BEC \cite{liberati}. 
\subparagraph{Appendix B: Inertial and accelerated motion}
Having a description of the BEC on a spacetime metric enables us to describe it while it undergoes inertial and uniformly accelerated motion. In the inertial case, we consider Minkowski coordinates $(t,x)$ where the line element is given by 
$ds^2=g_{\mu\nu}dx^{\mu}dx^{\nu}=-c^2dt^2+dx^2$.  Considering that the spacetime metric $g_{ab}$ is flat, we find from inspection of Eq. (\ref{metric}) that the effective metric is also flat when the spatial flow velocities vanish. In this case the phonons obey a Klein-Gordon equation which takes the form of a wave equation in Minkowski coordinates with propagation velocity $c_s$.  The solutions to the equation, denoted $\phi_n(t,x)$ with $n\in\mathbb{N}$, form an orthonormal set of modes in terms of which the field $\Pi(t,x)$ can be expanded,
\begin{eqnarray}
\Pi(t,x)=\sum_n\left[\phi_n(t,x)a_n+\text{h.c.}\right].
\end{eqnarray}
Here $a_n,a^{\dagger}_n$ are the annihilation and creation operators associated to the modes $\phi_n$.
 For periods of uniform acceleration, Rindler coordinates $(\eta,\chi)$  are a convenient choice of coordinates \cite{birrelldavies}. They are related to the Minkowski coordinates by the following transformation
\begin{eqnarray}
t&=&\frac{\chi}{c_s}\sinh \eta\nonumber\\
x&=&\chi\cosh \eta,\label{right:rindler:wedge}
\end{eqnarray}
where $\chi>0$ has dimension length and $\eta\in\mathbb{R}$ is the dimensionless Rindler time. The line element in these coordinates is $ds^2=-\chi^2d\eta^2+d\chi^2$. A uniformly accelerated observer follows a trajectory of constant $\chi=\chi_o$ and its proper time is given by $\tau=\frac{c_s}{a}\eta$, where $a=\frac{c_s^2}{\chi_o}$ is its proper acceleration. When the BEC undergoes acceleration, the background density can become inhomogeneous. In this case, it is not possible to neglect the generalized kinetic operator and it is not possible to describe the condensate using quantum field theory in a curved spacetime. In that case, the field equation is given by Eq. (\ref{almost:final}). Fortunately, in the acceleration regimes we consider these effects are negligible. Indeed, mimicking the acceleration by an external potential of the form $V(x)= m\,\cdot a\, \cdot x$ \cite{acceleratedbec}, we obtain that the term associated to the quantum pressure $T_{\rho}$ can be safely neglected as long as $\frac{\partial_x\,\rho}{\rho}\simeq\frac{h}{L}\ll{\frac{m\,c}{\hbar}}$. Using the values of $h$ and $L$ that we considered in the main text, $h/L\simeq 10^3 \operatorname{m}^{-1}$ while $m\, c/\hbar$ is larger than $10^{15}\operatorname{m}^{-1}$. Therefore, when the BEC undergoes uniform acceleration, the phononic BEC field obeys again a Klein-Gordon equation which takes the form in this case of a wave equation in Rindler coordinates. The Rindler solutions are denoted by $\tilde{\phi}_n(\eta,\chi)$ with $n\in\mathbb{N}$ and the field expansion is given by
\begin{eqnarray}
\Pi(\eta,\chi)=\sum_n\left[\tilde{\phi}_n(\eta,\chi)\tilde{a}_n+\text{h.c.}\right].
\end{eqnarray}
The operators  $\tilde a_n,\tilde a^{\dagger}_n$ are now the annihilation and creation operators associated to the Rindler modes $\tilde\phi_n$.
The effects of the inhomogeneity cannot be addressed with the mathematical formalism used in our analysis. Preliminary results addressing this point using numerical methods show that the effects are indeed small and can be neglected. Such numerical analysis will be published elsewhere. Since in the context of a quench of the BEC \cite{davidnicoivy} the inhomogeneity produces mode mixing between modes other than k, k', we anticipate that larger inhomogeneity will produce further entanglement degradation in our system.
\subparagraph{Appendix C: Bogoliubov transformations, the covariance matrix formalism and entanglement}

In our work we consider a condensate which is initially inertial, undergoes a change in the gravitational field strength as it changes into a different orbit and is finally inertial again. The change in field strength corresponds to a period of uniform acceleration. The mode creation and annihilation operators in the initial and final regions denoted by $a,a^{\dagger}$ and $\hat a, \hat a^{\dagger}$ respectively, are related through a Bogoliubov transformation \cite{birrelldavies},
\begin{align}
\begin{pmatrix}
\hat{a}\\
\hat{a}^{\dagger}
\end{pmatrix}
=
\begin{pmatrix}
\alpha & \beta\\
\beta^* & \alpha^* 
\end{pmatrix}
\cdot
\begin{pmatrix}
a\\
a^{\dagger}
\end{pmatrix},\label{total:bogo:transformation}
\end{align}
where $\alpha_{nm}=(\phi_n,\hat\phi_m)$ and $\beta_{nm}=-(\phi_n,\hat\phi_m^{\ast})$ are Bogoliubov coefficients. Here $(\cdot,\cdot)$ denotes the inner product. $\phi$ and $\hat\phi$ are Minkowski mode solutions in the initial and final regions, respectively.  These Bogoliubov coefficients are functions of the Bogoliubov coefficients between the Rindler and Minkowski modes  given by  ${}_0\alpha_{nm}=(\phi_n,\tilde\phi_m)$ and ${}_0\beta_{nm}=-(\phi_n,\tilde\phi_m^{\ast})$ and of phases acquired during the period of uniform acceleration where the condensate undergoes free evolution (for more details see \cite{alphacentauri}). When $h=aL/c_s^2\ll1$ is a small, it is possible to expand the Bogoliubov coefficients \eqref{total:bogo:transformation} in series as
\begin{eqnarray}
\alpha_{mn}&=&\alpha^{(0)}_{mn}+\alpha^{(1)}_{mn}+\alpha^{(2)}_{mn}+\mathcal{O}(h^3)\nonumber\\
\beta_{mn}&=&\beta^{(1)}_{mn}+\beta^{(2)}_{mn}+\mathcal{O}(h^3),\label{bogo:coefficient:expansion}
\end{eqnarray}
where the superscript ${}^{(n)}$ denotes quantities that are proportional to $h^n$ \cite{alphacentauri,Friis:Lee:Bruschi:Louko:12}. In the case we consider here, the Bogoliubov coefficients to first order in $h$ are given by \cite{alphacentauri,Friis:Lee:Bruschi:Louko:12} ,
\begin{eqnarray}\label{bogos}
\alpha^{(0)}_{mn}&=&\delta_{mn}e^{-i\Omega_n\Delta\tau}\nonumber\\
\beta^{(0)}_{mn}&=&0\nonumber\\
\alpha^{(1)}_{mn}&=&e^{-i(\Omega_n-\Omega_m)\Delta\tau}\,{}_0\alpha^{(1)}_{mn}=\nonumber\\& &e^{-i(\Omega_n-\Omega_m)\Delta\tau}\,\frac{(- 1+(-1)^{(m-n)}) \sqrt{m\,n}}{\pi^2\, (m-n)^3}\nonumber\\
\beta^{(1)}_{mn}&=&e^{i(\Omega_n-\Omega_m)\Delta\tau}\,{}_0\beta^{(1)}_{mn}=\nonumber\\& &e^{i(\Omega_n-\Omega_m)\Delta\tau}\,\frac{(1-(-1)^{m-n}) \sqrt{m\,n}}{\pi^2\, (m+n)^3}.
\end{eqnarray}
where $\Omega_n$ are the frequencies of the modes as measured by a comoving accelerated observer and $\Delta\tau$ is the proper time spent while accelerating. 

Let us now consider the covariance matrix formalism, in which all the relevant information about the state is encoded in the first and second moments of the field. In particular, the second moments are described by the covariance matrix $\sigma_{ij}=\langle X_{i} X_{j}+X_{j}X_{i}\rangle-2\langle X_{i}\rangle\langle X_{j}\rangle$, where $\langle\,.\,\rangle$ denotes the expectation value and the quadrature operators $X_{i}$ are the generalized position and momentum operators of the field modes given by $X_{2n-1}=\frac{1}{\sqrt{2}}(a_{n}+a^{\dag}_{n})$ and $X_{2n}=\frac{-i}{\sqrt{2}}(a_{n}-a^{\dag}_{n})$.  Every unitary transformation in Hilbert space that is generated by a quadratic Hamiltonian can be represented as a symplectic matrix $S$ in phase space. These transformations form the real symplectic group $Sp(2n,\mathds{R})$, the group of real $(2n\times2n)$ matrices that leave the symplectic form $\Omega$ invariant, i.e., $S\Omega S^{T}=\Omega$, where 
$\Omega=\bigoplus_{i=1}^{n}\Omega_i$ and $\Omega_i=\left(
            \begin{array}{cc}
              0 & 1 \\
              -1 & 0 \\
            \end{array}
          \right)$\,. The time evolution of the field, as well as the Bogoliubov transformations, can be encoded in this symplectic structure (for details see\cite{nicoivy}).  The covariance matrix after a symplectic transformation is given by $\tilde{\sigma}=S\sigma S^{T}$. In our proposal Valentina and Yuri are initially inertial and prepare an entangled two-mode squeezed state of their phononic modes $k$ and $k^{\prime}$, each one of them in their respective condensate.  We assume that all other modes in both condensates are in the vacuum state. Using that the trace operation over a set of modes is implemented in this formalism by deleting the rows and columns associated to those modes, we find that the covariance matrix of the reduced state for the modes $k$ and $k^{\prime}$ is given by,
\begin{eqnarray}
    \sigma_{kk^{\prime}}\,&=&\,
        \begin{pmatrix}
            \cosh(2r) \mathds{1}_{2}     &   \phi_{kk^{\prime}}    \\
            \phi_{kk^{\prime}}          &   \cosh(2r)  \mathds{1}_{2}  \\
        \end{pmatrix}\ \mbox{where}\
       \nonumber\\ \phi_{kk^{\prime}}\,&=&\, \begin{pmatrix}
                                        \sinh(2r)    &   0           \\
                                        0           &   -\sinh(2r)   \\
                                \end{pmatrix}\,,
    \label{eq:two mode squeezed cov matrix}
\end{eqnarray}
and  $r$ is the squeezing parameter of the state. The matrix $\mathds{1}_{2}$ is the $2\times 2$ identity matrix. The covariance matrix after Valentina remains inertial and Yuri undergoes a single period of uniform acceleration to move to a different orbit is given by 
\begin{align}
    \tilde{\sigma}_{k,k^{\prime}}\,=\,
        \begin{pmatrix}
            C_{kk}               &   C_{kk^{\prime}} \\
            C_{kk^{\prime}}^{T} &   C_{k^{\prime}k^{\prime}}  \\
        \end{pmatrix}\,,
    \label{eq:transformed two mode squeezed cov matrix}
\end{align}
where $C_{kk}=\cosh(2r)\mathds{1}_{2}$, $C_{kk^{\prime}}=\phi_{kk^{\prime}}\mathcal{M}^{T}_{k^{\prime}k^{\prime}}$ and
\begin{align}
    C_{k^{\prime}k^{\prime}}\,=\,\cosh(2r)\mathcal{M}_{k^{\prime}k^{\prime}}\mathcal{M}^{T}_{k^{\prime}k^{\prime}}\,+\,
                        \sum\limits_{n\neq k^{\prime}}\mathcal{M}_{k^{\prime}n}\mathcal{M}^{T}_{k^{\prime}n}\,.
    \label{eq:mode k prime diagonal block}
\end{align}
The  $2\times2$ matrices $\mathcal{M}$ encode the Bogoliubov coeffcients  given by Eq. (\ref{bogos}),
\be\label{Mmatrices}
\mathcal{M}_{nm}=\left(
                   \begin{array}{cc}
                     \mathrm{Re}(\alpha^{(0)}_{mn}+\alpha^{(1)}_{mn}-\beta^{(1)}_{mn}) & \mathrm{Im}(\alpha^{(0)}_{mn}+\alpha^{(1)}_{mn}+\beta^{(1)}_{mn}) \\
                     -\mathrm{Im}(\alpha^{(0)}_{mn}+\alpha^{(1)}_{mn}-\beta^{(1)}_{mn}) & \mathrm{Re}(\alpha^{(0)}_{mn}+\alpha^{(1)}_{mn}+\beta^{(1)}_{mn})
                   \end{array}
                 \right)\,.
\ee
 Here $\mathrm{Re}$ and $\mathrm{Im}$ denote the real and imaginary parts, respectively.  A number of computable measures of entanglement exist for Gaussian states in terms of the smallest symplectic eigenvalue $\nu_{-}$ of the  partial transposition of $\tilde{\sigma}$. Here we are interested in computing the negativity of the state $\tilde{\sigma}_{kk^{\prime}}$ to understand how entanglement is affected when Yuri has changed his condensate into an orbit with different gravitational potential. In this case the negativity is given by
\begin{equation}
N=\operatorname{max}[0,\frac{1-\nu_{-}}{2\,\nu_{-}}] \label{eq:negsym}
\end{equation}
where
\begin{align}
    \nu_{\pm}\,=\,\sqrt{\frac{\Delta(\tilde{\sigma}_{kk^{\prime}})\pm\sqrt{\Delta^{2}(\tilde{\sigma}_{kk^{\prime}})-4 \det\tilde{\sigma}_{kk^{\prime}}}}{2}}\,
    \label{eq:nutildeminus}
\end{align}
and $\Delta(\tilde{\sigma}_{kk^{\prime}})=\det C_{kk}+\det C_{k^{\prime}k^{\prime}}-2 \det C_{kk^{\prime}}$. Using Eqs. (\ref{bogo:coefficient:expansion}) to (\ref{eq:nutildeminus}) we obtain our main result which is given by Eq. (5) in the main text  and
\begin{equation}
f^{\alpha}_{k'}=\sum_n |\alpha^{(1)}_{k'n}|^2 ,  f^{\beta}_{k'}=\sum_n |\beta^{(1)}_{k'n}|^2 \label{sums}.
\end{equation}

\section*{Acknowledgements} 
The authors would like to acknowledge Stefano Liberati and Tim Ralph for useful discussions and comments. C. S and I. F. acknowledge support from EPSRC (CAF Grant No. EP/G00496X/2 to I. F.) A. W. acknowledges funding from EPSRC grant No. EP/H027777/1. VB acknowledges support by a Victoria University PhD scholarship. D. K. L. O acknowledges support from QUISCO. V. B. and D. E. B. would like to thank the hospitality of the University of Nottingham.

\end{document}